\documentclass[aps,prl,twocolumn,superscriptaddress,showpacs,floatfix]{revtex4-1}
\usepackage{mathrsfs}
\usepackage[figuresright]{rotating}
\usepackage{amsmath}
\usepackage{amssymb}
\usepackage{graphicx}
\usepackage{color}
\usepackage{dcolumn}
\usepackage{bm}
\usepackage{siunitx}

\usepackage{soul}

\makeatletter

\newcommand{\Rmnum}[1]{\expandafter\@slowromancap\romannumeral #1@}
\makeatother

\begin{document}

	\title{Giant Anomalous Hall Effect in Kagome Nodal Surface Semimetal Fe$_3$Ge}

	\author{Shu-Xiang Li$^*$}
	\affiliation{School of Physics, Zhejiang University, Hangzhou, 310027, China}
	
	\affiliation{State Key Laboratory of Silicon and Advanced Semiconductor Materials, Zhejiang University, Hangzhou, 310027, China}
	
	\author{Wencheng Wang$^*$}
	
	\affiliation{National Laboratory of Solid State Microstructures and School of Physics, Nanjing University, Nanjing 210093, China
		and Collaborative Innovation Center of Advanced Microstructures, Nanjing University, Nanjing 210093, China}
	
	\affiliation{International Quantum Academy, Shenzhen 518048, China}
	
	\author{Sheng Xu}
	\affiliation{School of Physics, Zhejiang University, Hangzhou, 310027, China}
	
	\affiliation{Department of Applied Physics, Zhejiang University of Technology, Hangzhou 310023, China}

	\author{Tianhao Li}
	\affiliation{School of Physics, Zhejiang University, Hangzhou, 310027, China}
	
	\affiliation{State Key Laboratory of Silicon and Advanced Semiconductor Materials, Zhejiang University, Hangzhou, 310027, China}
	
	\author{Zheng Li}
	\affiliation{School of Physics, Zhejiang University, Hangzhou, 310027, China}
	
	\affiliation{State Key Laboratory of Silicon and Advanced Semiconductor Materials, Zhejiang University, Hangzhou, 310027, China}
	
	\author{Jinjin Wang}
	\affiliation{School of Physics, Zhejiang University, Hangzhou, 310027, China}
	
	\affiliation{State Key Laboratory of Silicon and Advanced Semiconductor Materials, Zhejiang University, Hangzhou, 310027, China}
	
	\author{Jun-Jian Mi}
	\affiliation{School of Physics, Zhejiang University, Hangzhou, 310027, China}
	
	\affiliation{State Key Laboratory of Silicon and Advanced Semiconductor Materials, Zhejiang University, Hangzhou, 310027, China}
	
	\author{Qian Tao}
	\affiliation{School of Physics, Zhejiang University, Hangzhou, 310027, China}

	\author{Feng Tang}
	\affiliation{National Laboratory of Solid State Microstructures and School of Physics, Nanjing University, Nanjing 210093, China
		and Collaborative Innovation Center of Advanced Microstructures, Nanjing University, Nanjing 210093, China}
	
	\author{Xiangang Wan}
	\affiliation{National Laboratory of Solid State Microstructures and School of Physics, Nanjing University, Nanjing 210093, China
		and Collaborative Innovation Center of Advanced Microstructures, Nanjing University, Nanjing 210093, China}
	
	\affiliation{Hefei National Laboratory, Hefei 230088, China}
	
	
	\author{Zhu-An Xu}
	
	\affiliation{School of Physics, Zhejiang University, Hangzhou, 310027, China}
	
	\affiliation{National Laboratory of Solid State Microstructures and School of Physics, Nanjing University, Nanjing 210093, China
		and Collaborative Innovation Center of Advanced Microstructures, Nanjing University, Nanjing 210093, China}
	
	\affiliation{Hefei National Laboratory, Hefei 230088, China}
	
	\email{zhuan@zju.edu.cn }
	
	

	\begin{abstract}
		It is well known that the intrinsic anomalous Hall effect (AHE) arises from the integration of the non-zero Berry curvature (BC), conventionally observed in the Dirac/Weyl and nodal-line semimetals. Moreover, nodal surface semimetals are expected to exhibit more significant BC under the prevalence of degenerate points near the Fermi level. In this work, we report the detection of a giant AHE in the Kagome magnet Fe$_3$Ge with a two-dimensional (2D) nodal surface (NS) at $k_{z}=\pi$ plane, exhibiting an anomalous Hall conductivity (AHC) of 1500 $\Omega^{-1}$cm$^{-1}$ at 160 K, the highest among all reported Kagome topological materials. This finding suggests a new platform for searching large AHC
		materials and facilitates potential room-temperature applications
		in spintronic devices and quantum computing.
		
	\end{abstract}
	
	\maketitle
	
	For the progress of cutting-edge technological apparatuses, the identification and synthesis of materials exhibiting extraordinary physical characteristics are of paramount importance \cite{1-N2005graphe,2-N2005,3NbP2015,4QH2006,5Cd3As2,16TE,17TE}. In recent years, the interplay between magnetism and topology in magnetic topological materials (MTMs) has attracted great interest and triggered the upsurge of frontier research \cite{6GdPtBi,7Co2TiX,8Weyl,7berry-phase,8sph}. The anomalous Hall effect (AHE) is a characteristic phenomenon in MTM that has substantial potential for application in spintronic devices \cite{9spintronic,12AHE}, which is ascribed to the existence of non-zero Berry curvature (BC) that arises from the band crossing and can be described as \cite{15AHEsigma}
	\begin{equation}\label{equ1}
		\sigma_{\alpha \beta }(T,\mu )=-\epsilon _{\alpha \beta \gamma } \frac{e^{2}}{\hbar}{\int}\frac{d \boldsymbol{k} }{2\pi ^{3}}\Omega_{n,\gamma }(\boldsymbol{k})f_{n,k}
	\end{equation}
	where $f_{n,k}$, $\mu$, $\epsilon _{\alpha \beta \gamma }$, and $\Omega_{n,\gamma }$ are the Fermi-Dirac distribution function with the band index $n$ and the wave vector $k$, the chemical potential, Levi-Civita symbol, and the $\gamma$ component of the BC, respectively.
	The magnitude of AHC is intricately correlated with the nature of the band crossing and the Fermi energy relative to the band crossing point \cite{12Mn3Sn,2CoSnS-2018,TbMnSn6}.

	In time-reversal symmetry broken topological systems, the band crossings are isolated zero-dimensional (0D) points near the Fermi level.
	The typical magnetic Weyl semimetal such as Mn$_3$Sn, Co$_3$Sn$_2$S$_2$ and
	RMn$_6$Sn$_6$ (R= Lanthanide rare earths) are reported to exhibit the intrinsic AHC in the order of
	10$^2$ $\Omega^{-1}$cm$^{-1}$  \cite{12Mn3Sn,2CoSnS-2018,TbMnSn6}. The massive Dirac material Fe$_3$Sn$_2$ is suggested to hold an intrisic AHC of 200 $\Omega^{-1}$cm$^{-1}$ \cite{7massive-dirac}. In addition to the touching of bands at a node, they may also form a one-dimensional (1D) line. In nodal-line ferromagnets MnAlGe, Co$_2$MnAl, the intrinsic AHC is enhanced to about 700, 1600 $\Omega^{-1}$cm$^{-1}$, respectively
	\cite{9MnAlGe,11co2mnal}.
	
	Furthermore, there is a remaining possibility to form a three-dimensional(2D) nodal surface (NS) in three-dimensional (3D) Brillouin zone (BZ), each point on the surface is doubly degenerate with the protection of non-symmorphic symmetries \cite{7NS-PRB2018}. Such a nodal-surface system expect to be a promising platform for exploring the anomalous transport phenomena, which maximally improve the effective energy integral paths \cite{10MnSi-nature}. However, reports on their transport properties and electronic structures remain relatively scarce in magnetic materials that are constrained by the unique magnetic symmetry protection \cite{8ZrSiS,Ti3Al,BaVS}.

	In this letter, we report a ferromagnetic nodal-surface Kagome compound Fe$_3$Ge with a giant AHC of 1500 $\Omega^{-1}$cm$^{-1}$ at 160 K for the $B$//$ab$ configuration, the highest among the reported Kagome topological materials.
	The room temperature anomalous Nernst thermopower of about 2.8 $\mu$VK$^{-1}$ is also detected. Theoretical analysis reveals that there is a nontrivial NS at $k_{z}=\pi$ plane near the Fermi level, which is protected by the combination of two-fold screw rotational symmetry and time-reversal symmetry. It remains robust when considering the spin-orbit coupling (SOC). First-principles calculations indicate that the BC in Fe$_3$Ge is induced by the NS bands and prominently enhanced at the intersections of the NS and the Fermi surface (FS). It provides a clearer and more detailed understanding of the observed enhancement of the AHC in the nodal surface materials. Our study suggests a novel experimental platform for searching large AHC materials and facilitates room-temperature potential applications in spintronic devices and quantum computing, within a rarely explored topological phase.

	\begin{figure}[htbp]
		\centering
		\includegraphics[width=0.5\textwidth]{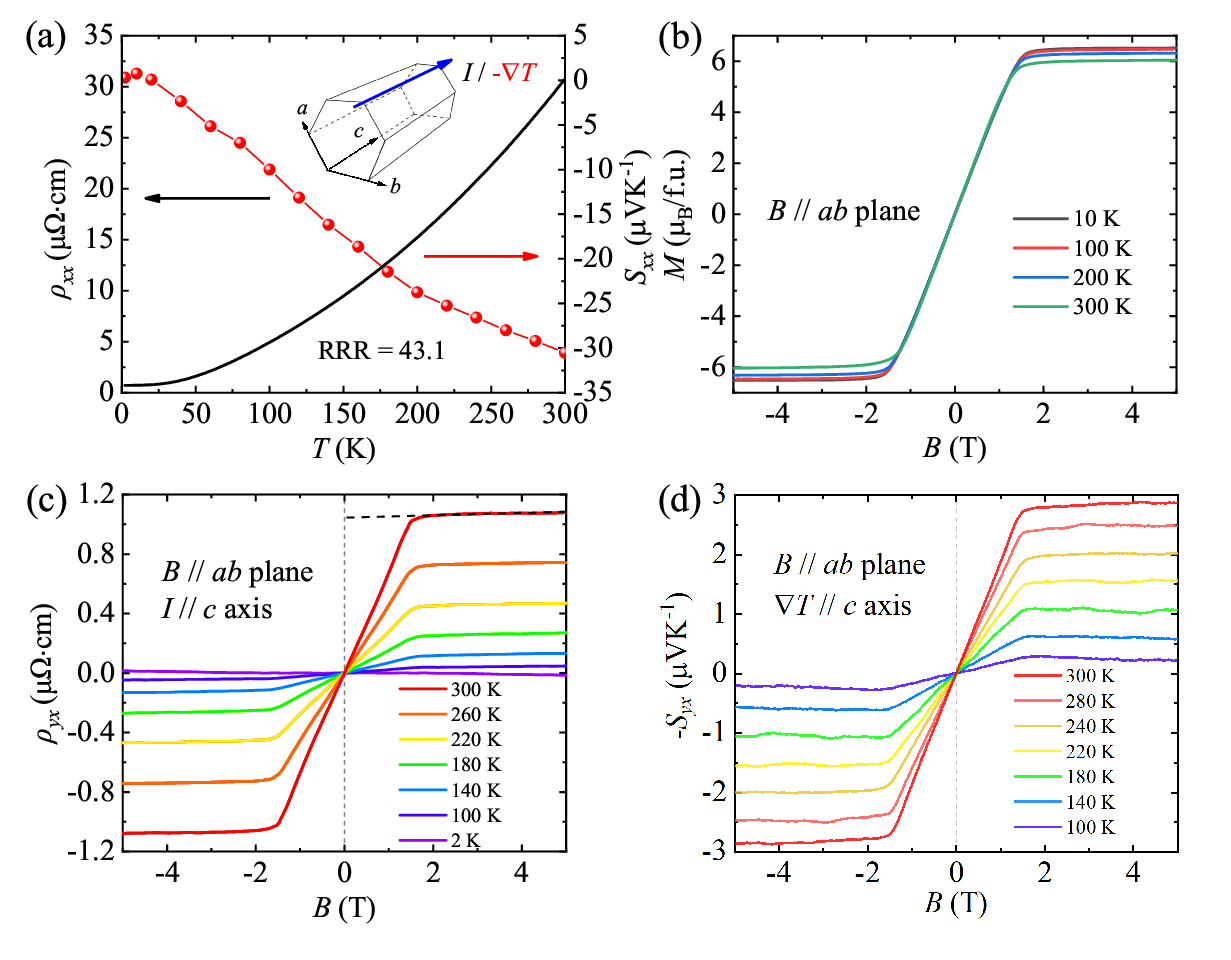}\\
		\caption{(a) Temperature dependence of longitudinal resistivity (black) and Seebeck coefficient (red).  (b) Magnetic field dependence of magnetization $M$ ($\mu_B$/f.u.) with $B//ab$ plane. (c) Hall resistivity as a function of magnetic field when $B//ab$ plane and $I//c$ axis. The black dash line is a linear fitting of Hall resistivity at high field when $T$ = 300 K, which intercept is $\rho_{yx}^{A}$. (d) Magnetic field dependence of Nernst coefficients when $B//ab$ plane, $\nabla T$ along the $c$ axis.}\label{2}
	\end{figure}
	
	Fe$_3$Ge crystallizes in a DO$_{19}$ structure with the space group of P6$_3$/mmc \cite{44transiton-fe3ge}, $a$ = 5.175 {\AA} and $c$ = 4.226 {\AA}. Its crystal growth details can be found in the Supplementary
	Materials. Comparing with its sibling compounds FeGe \cite{42FeGe} and Fe$_3$GeTe$_2$ \cite{10Fe3GeTe2}, the crystal structure of Fe$_3$Ge consists only of the directly stacked Fe-Ge Kagome layers, as shown in Figs. S2(a) and (b) in the Supplementary Materials. The primitive unit cell is composed of two Kagome layers with a 60$^{\circ}$ rotation, where the magnetic Fe atoms are arranged on a Kagome lattice in the $ab$ plane and the Ge atoms occupy the center of each hexagon.
	Fig. 1(a) exhibits the zero-field longitudinal resistivity $\rho_{xx}$ and the Seebeck coefficient as functions of temperature, with the current and heat flow along the $c$ axis, respectively. The $\rho_{xx}$ decreases with the temperature decreasing, exhibiting an extremely low residual resistivity of about 0.7 $\mu\Omega\cdot$cm at 2 K and a high residual resistivity ratio ($\rho _{xx}(300k)/\rho _{xx}(2k)$) of 43.1, indicating the metallic behavior and the high quality of the sample.

	Previous neutron diffraction measurements indicate that the Curie temperature ($T_C$) is about 629 K and a spin-flip transition occurs around 382 K \cite{40Fe3Ge-neutron2023}. The total moments transit from $c$-axis in $ab$ plane below 382 K, 
	which demonstrated in temperature-dependent resistivity results under 400 K, as shown in Fig. S3(a) (in the Supplementary Materials). The detailed magnetic measurements, as shown in Fig. 1(b) and Fig. S7 (in the Supplementary Materials), exhibit the characteristics of soft ferromagnetism, consistent with previous reports \cite{,44transiton-fe3ge,39Fe3Ge-magnetic-1976,43Si-Fe3Ge-2015,45GaAs-Fe3Ge}.

	Fe$_3$Ge is isostructural to the M$_3$X compounds (M = Mn, Fe; X = Sn, Ge, Ga), which are featured with Kagome lattice, exhibiting abundant topological properties and excellent anomalous transport performance \cite{28Fe3Sn,12Mn3Sn,13Mn3Ge,36Mn3Ga,48Fe3Ga-al}. We measured the Hall resistivity in Fe$_3$Ge with the configuration of $B//ab$, schematic measurements setup as shown in Fig. S1. Fig. 1(c) represents $\rho _{yx}$ as a function of field at various temperatures. It increases rapidly in the low magnetic field regime, followed by saturation upon further increase of the magnetic field.
	The total Hall conductivity ($\sigma _{xy}\approx\rho_{yx}/\rho_{xx}^2$) is presented in Fig. 2(a) \cite{15AHEsigma}. The high-field linear Hall effect is projected towards zero magnetic field to obtain the $\rho _{yx}^{A}$ as the black dashed line in Fig. 1(c). Following a same process, the AHC at various temperatures are obtained, as shown in Fig. S5(a). A maximum $\sigma _{xy}^{A}$ of about 1500 $\Omega^{-1}$cm$^{-1}$ was measured at 160 K, 
	which is higher than that of many typical Kagome materials, such as Mn$_3$Sn, Fe$_3$Sn$_2$, Fe$_3$GeTe$_2$ and Co$_3$Sn$_2$S$_2$ \cite{12Mn3Sn,4Fe3Sn2-prb,10Fe3GeTe2,2CoSnS-2018}. When normalized to the number of contributing atomic planes, the AHC is $\approx$ 0.82 $\frac{e^2}{h}$ per Fe layer.

	The saturation value of $\rho _{yx}^A$ exhibits a decreasing trend with the reduction in temperature, which progressively diminishing below 80 K and nearly disappearing at 2 K, as illustrated in Fig. 1(c). This behavior diverges from the behavior observed in the polycrystals \cite{41Fe3Ge-apl}, indicating that the absence of grain boundary effects may impact a significant influence. A two-band model is applied to describe the carrier property of Fe$_3$Ge in 2 K (see Fig. S3(b) and Table. S1 in the Supplementary Materials). The extracted carrier concentration of electron and hole are $n_e\sim $ 1.2 $\times$ 10$^{21}$\ cm$^{-3}$ and $n_h\sim $ 1.18 $\times$ 10$^{21}\ $cm$^{-3}$ respectively.
	
	In order to understand the origin of AHE in Fe$_3$Ge, we scale the experimental anomalous Hall resistivity data with the longitudinal resistivity using the equation \cite{47fitting} :
	\begin{equation}\label{equ2}
		\rho_{yx}^{A}(T)=(\alpha\rho_{xx0}+\beta\rho_{xx0}^2)+\gamma\rho_{xx}^2(T)
	\end{equation}
	where $\rho_{xx0}$ is residual longitudinal resistivity. $\alpha$, $\beta$, $\gamma$ are from the contribution of skew scattering, side-jump and intrinsic BC, respectively \cite{19side-jump,18IAHE,20luttinger}. In Fig. 2(b), the plot of $\rho_{yx}^{A}$ versus $\rho_{xx}^2$ exhibits an almost linear behavior between 80 K and 300 K, the slope $\gamma$ provides an estimation of the intrinsic AHC. Some data slightly deviate from the fitting line. The origin of such deviation is not very clear and also observed in other compounds \cite{2CoSnS-2018,4Fe3Sn2-prb,9MnAlGe,TbMnSn6}. In Fe$_3$Ge, since the topological states in the topological materials are concentrated near the Fermi energy ($E_F$), a slight shift in $E_F$ will significantly influence the AHC, leading to a sharp peak near $E_F$, as evident in Fig. 3(d). The rough linear fit of the $\rho_{yx}^{A}$ versus $\rho_{xx}^2$ from 80 K to 300 K gives an intrinsic AHC of 1138 $\Omega^{-1}$cm$^{-1}$, and confirms that the giant AHE in Fe$_3$Ge is primarily due to the intrinsic mechanism.
	
	\begin{figure}[htbp]
		\centering
		\includegraphics[width=0.50\textwidth]{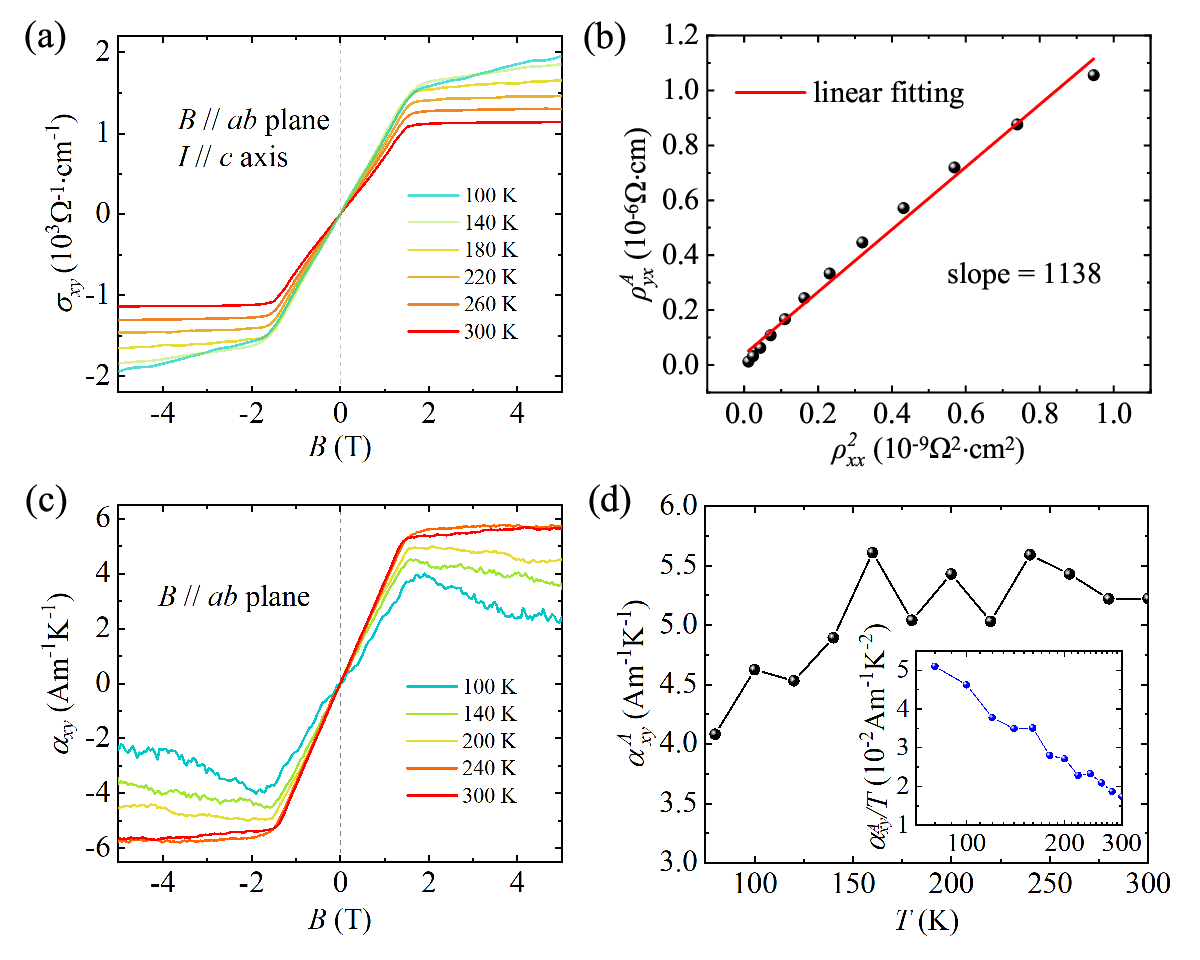}\\
		\caption{(a) Hall conductivity as a function of magnetic field for the configuration of $B//ab$ plane and $I//c$ axis. (b) Scaling plot of $\rho_{yx}^{A}$ with $\rho_{xx}^2$. A linear fitting (red line) is applied from 80 K to 300 K, where the slope represents the intrinsic anomalous Hall conductivity. (c) Magnetic field dependence of transverse thermoelectric conductivity. (d) Temperature dependence of ATC. Inset: $\alpha_{yx}/T$ versus $T$ with logarithmic scale. }\label{2}
	\end{figure}
	
	\begin{figure*}[htbp]
		\centering
		\includegraphics[width=0.9\textwidth]{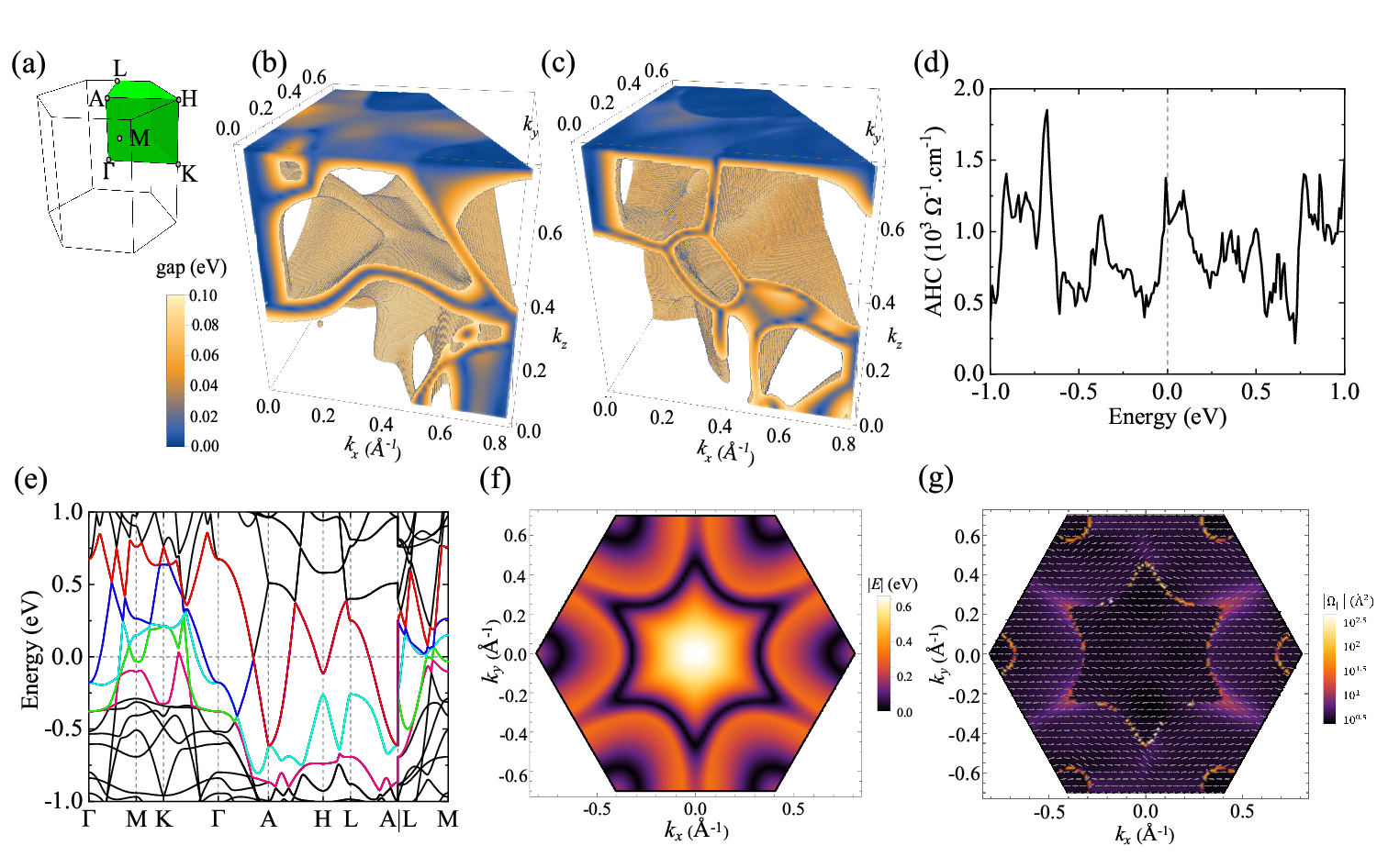}\\
		\caption{ (a) 3D first BZ with high symmetry points. Green section shows the irreducible BZ. (b) (c) Band gap $\triangle$$E$ (eV)  between the 51st (green) and 52st (cyan) bands, 53st (blue) and 54st (red) bands, respectively, within the irreducible BZ. We selectively display the band gap at $\triangle$$E <$ 0.1 eV. (d) Energy dependence of AHC with the magnetic moment along the $a$-axis. (e) Band structure with the direction of magnetic moment along the $a$-axis. SOC is considered and $U$ = 1 eV. There are five bands crossing the Fermi level, and band degeneracy is observed along the A-L-H-A path. (f) Energy distribution at $k_z=\pi$ plane. The color bar shows the energy relative to $E_F$. (g) Contour plot of the BC distribution at $k_z=\pi$ plane. $\left | \Omega_{\left |  \right | }  \right | $ is the sum of BC over the occupied states, and the gray arrow indicates the direction of BC.
		}\label{2}
	\end{figure*}
	
	Following the demonstration of giant AHC in Fe$_3$Ge, we proceed to investigate the Nernst thermopower ($S_{xy}$). In contrast to Hall conductivity, which accounted for all occupied electronic states, Nernst thermopower is sensitive to the electronic states near the Fermi energy, making it a crucial tool for studying Fe$_3$Ge wherein the topological features exist near the $E_F$. Fig. 1(d) shows the field-dependent S$ _{xy}$ at different temperatures, corroborating its anomalous behavior consistent with the Hall resistivity and magnetization profile, as shown in Figs. 1(b) and 1(c). A large value of $\left |S _{yx}^{A}  \right |$ reaches 2.8 $\mu$VK$^{-1}$ at 300 K, which significantly exceeds the value anticipated from the trivial ferromagnets. For conventional ferromagnets, the $\left |S _{yx}^{A}  \right |$ is proportional to the magnetization with the relation of $\left |S _{yx}^{A}  \right | =\left | Q_{s}\right | \mu_{0} M$, where $\left | Q_{s}\right | $ is the anomalous Nernst coefficient, ranging from 0.05 to 1 $\mu$VK$^{-1}$T$^{-1}$ \cite{Qs}. According to this relationship, Fe$_3$Ge exhibits a maximum $\left |S _{yx}^{A}  \right |$ = 1.4 $\mu$VK$^{-1}$T$^{-1}$, which is lower than the observed value at 300 K. This connects the ANE in Fe$_3$Ge to nontrivial mechanism.

	Fig. 2(c) presents the transverse thermoelectric conductivity $\alpha_{xy}$ versus field using the following equation, $\alpha_{xy}= S_{xy}\sigma_{xx}+\sigma _{xy}S_{xx}$, where $\sigma_{xx} = 1/\rho_{xx}$ and $S_{xx}$ are the longitudinal conductivity and Seebeck coefficient, respectively. In Fig. 2(d), we plot the $\alpha_{xy}^A$ as a function of $T$, where the $\alpha_{xy}^A$ is the intercept obtained from linear fitting of $\alpha_{xy}(B)$ at saturation field. A maximum value of approximately 5.6 A$m^{-1}$K$^{-1}$ is observed. At low temperature ($k_{B}T\ll E_{F}$), the $\alpha_{yx}$ usually exhibit a T-linear behavior, follow the Mott relation \cite{14mott}, $\alpha_{yx}=-\frac{\pi^{2}}{3}\frac{k_{B}^{2}T}{|e|}(\frac{\partial\sigma_{yx}}{\partial E})_{E=E_{F}}$. However, within the temperature range we measured (80 K - 300 K), the behavior of $\left | \alpha_{yx}^A\right | \sim -Tlog(T)$ displays a violation of the Mott relation. It could be induced by the topological states of NS, where shifts in the chemical potential modulate the BC. A similar behavior has been observed in other materials, though a unified explanation has yet to be established \cite{28Fe3Sn,48Fe3Ga-al,8co2mnga-nst}.

	Thereafter, the first-principles calculations are applied to
	investigate the possible origin of intrinsic anomalous response in
	Fe$_3$Ge. The calculation details are presented in the
	Supplementary Materials. With the Fe moments along $\left [ 100
	\right ] $ direction, the symmetries of Fe$_3$Ge are reduced from
	the SG 194 (P6$_3$/mmc) to the magnetic SG 63.463 (Cm$^{\prime
	}$cm$^{\prime }$). Two-fold screw rotation symmetry is included in
	the magnetic SG along the $z$-direction, denoted as $S_{2z} : (x,
	y, z) \to (-x, -y, z + 0.5)$, it satisfies $T^2=1$, where $T$ is
	the time-reversal symmetry. Squaring the $S_{2z}$ operation on a
	Bloch state is equivalent to a translation along the $z$-direction
	by a lattice constant. Hence, the combination $S_{2z}$ and $T$
	satisfies $(TS_{2z})^2=e^{-ik_z} $, where $k_z$ is the wavevector
	in the $z$ direction. On the $k_{z}=\pi$ plane, $(TS_{2z})^2=-1$.
	By Kramers theorem \cite{51kramers}, two-fold degeneracy applies
	at every point on this plane, forming a topological NS that
	remains robust even in the presence of SOC.
	
	In the first-principles calculations, $U$ is introduced to account
	for the correlation effects of the Fe-$d$ electrons. The
	calculated carrier concentration aligns well with experimental
	data when $U$ = 1 eV (see Table S1 in the Supplementary
	Materials). The band structure with $U$ = 0 eV is presented in
	Fig. S6 in the Supplementary Materials, consistent with the recent
	report \cite{ruilou}. The high-symmetry points in the 3D BZ are
	depicted in Fig. 3(a). Two-fold degeneracy is observed in the
	A-H-L-A path at the $k{_z}=\pi$ plane as shown in Fig. 3(e). Away
	from this plane, the degeneracy is lifted due to the loss of
	symmetry protection. Notably, a previous study has reported the
	existence of a nodal plane structure in Fe$_3$Sn \cite{28Fe3Sn}.
	Nevertheless, this nodal plane fundamentally differs from the
	nodal surface discussed herein. In Fe$_3$Sn, the nodal plane arises from the crossing of spin-up and spin-down bands due to the absence of hybridization between these spin states, which is not protected by the crystal symmetry.
	
	To more intuitively exhibit the characteristics of the NS and gain
	a detailed understanding of its contribution to the BC in
	Fe$_3$Ge, we calculated the gap and BC distribution within the
	irreducible BZ (green sections of Fig. 3(a)). Fig. 3(e) shows five
	bands (51st to 55th) intersect the Fermi level. Specifically, the
	51st and 52nd bands, as well as the 53rd and 54th bands, form two
	sets of NSs. Figs. 3(b) and (c) exhibit the band gaps between the
	51st and 52nd, the 53rd and 54th bands, respectively, highlighting
	the band gaps where $\triangle E < 0.1$ eV. A notable zero gap is
	observed in the $k_{z}=\pi$ plane, which is attributed to the
	degeneracy in NS. Additionally, within the interior of the
	irreducible BZ, there are some gapless points due the accidental
	degeneracies. Considering the integration of BC over the occupied
	states, we theoretically calculated the Fermi-level-dependent AHC
	of Fe$_3$Ge in Fig. 3(d). A large AHC of about 1300
	$\Omega^{-1}$cm$^{-1}$ is observed near the Fermi level, which is
	in good agreement with experiments. Interestingly, the calculated
	AHC sustains a significant value across a broad range. Fig. 3(f)
	and (g) exhibit the energy $\left | E \right |$ and the BC
	distribution at $k_{z}=\pi$ plane, respectively. $\left |
	\Omega_{\left | \right | } \right |$ represents the BC component
	parallel to the $k_{z}=\pi$ plane, while the perpendicular
	component is constrained to zero by symmetry. The direction of the
	BC is indicated by the gray arrow. In the vicinity of zero energy
	in Fig. 3(g), where the NS intersects with the FS, substantial BC
	is generated, as shown in Fig. 3(g). It indicates that the BC
	arising from NS states contributes to the giant intrinsic AHC.
	
	For comparison, we measured AHE with another configuration of $B//c$ axis and $I//ab$ plane. The intrinsic AHC in experiment is 294 $\Omega^{-1}$cm$^{-1}$, significantly lower than the previous setup (see Fig. S8).
	Theoretical analysis indicates the magnetic SG is 194.270 (P6$_3$/mm$^{\prime }$c$^{\prime }$). In this configuration, the twofold screw rotation symmetry is broken, leading to the absence of topological nodal surface states and the lifting of band degeneracy along the H-L-A path, as shown in Fig. S6(c). A previous study reported the intrinsic BC of $B//c$ originates from the gapped nodal line structure \cite{41Fe3Ge-apl}. This suggests that the topological state of NS plays a crucial role in enhancing the AHC.

	Our investigations suggest a new system to search the large AHC compounds. It is well established that the magnitude of AHC is intricately associated with the degenerate points and their location relative to the Fermi level. Crystal symmetry enforce the crossing type, and the location is material-specific. The 2D crossing forming in NS systems is more extensive than that in Weyl/Dirac and nodal line systems. Additionally, the 2D NS usually exist at the BZ boundary, which at a Fermi level cut are generally 1D lines. It indicates that the NS structure will ensure that more crossing points are located near the Fermi level, enhancing the non-zero BC and the AHC significantly.

	
	In summary, a giant AHC of 1500 $\Omega^{-1}$cm$^{-1}$ and a large
	anomalous Nernst thermopower of 2.8 $\mu$VK$^{-1}$ are obtained in
	the Fe$_3$Ge single crystal. The crystal structure comprises a
	novel topological state of NS, which is responsible for the giant
	AHC. Its distinctive attributes, including the nodal-surface
	topology, large $\sigma _{xy}^{A}$/S$ _{xy}^{A}$, long-range
	in-plane ferromagnetic order with a high $T_c$ exceeding room
	temperature, and low production costs, making it a promising
	candidate for potential room temperature application in
	spintronics, quantum computing, and heat flow sensors. Our
	investigations deepen the comprehension of the accumulation in BC
	and suggest a new platform for searching compounds with large
	anomalous galvanomagnetic responses.
	
	S.Li and W.Wang contributed equally to this work. This work was
	supported by the National Natural Science Foundation of China
	(Grant No. 12174334, 12204410), the National Key \& Program of the
	China (Grant No. 2019YFA0308602), and the Innovation program for
	Quantum Science and Technology (Grant No. 2021ZD0302500).

\end{document}